\documentstyle[prd,floats,aps]{revtex}
\begin{document}
\title{Gauge invariant formalism for second order
perturbations of Schwarzschild spacetimes}
\author{Alcides Garat$^1$,
Richard H. Price$^2$\\
1. {\it Department of Physics, University of Utah, Salt Lake City, Utah
84112. On leave from Universidad de la Rep\'ublica, Montevideo, Uruguay.}\\
2. {\it Department of Physics, University of Utah, Salt Lake City, Utah
84112.}}

\maketitle
\begin{abstract}

The ``close limit,'' a method based on perturbations of Schwarzschild
spacetime, has proved to be a very useful tool for finding approximate
solutions to models of black hole collisions. Calculations carried
out with second order perturbation theory have been shown to give the
limits of applicability of the method without the need for comparison
with numerical relativity results. Those second order calculations
have been carried out in a fixed coordinate gauge, a method that
entails conceptual and computational difficulties. Here we demonstrate
a gauge invariant approach to such calculations. For a specific set
of models (requiring head on collisions and quadrupole dominance of
both the first and second order perturbations), we give a self
contained gauge invariant formalism. Specifically, we give (i) wave
equations and sources for first and second order gauge invariant wave
functions; (ii) the prescription for finding Cauchy data for those
equations from  initial values of the first and second fundamental
forms on an initial hypersurface; (iii) the formula for computing the
gravitational wave power from the evolved first and second order wave
functions.
\end{abstract}
\section{Introduction and overview}
\label{sec:intro}

In the next few years gravitational wave
antennas\cite{Ligo,Virgo,Geo,Tama} will go into operation with the
possibility of detecting astrophysical sources.  A plausible, and
certainly fascinating, origin of such waves 
would be the powerful burst of radiation generated in the merger of
two approximately equal mass black holes to form a single final
hole\cite{FlHu}.  A throrough understanding of this problem will
require numerical relativity on supercomputers, and is still several
years away. In the absence of numerical answers, some useful insights
have already been supplied by ``close limit'' perturbation
theory\cite{PP,APPSS,AP1,AP2,Bakeretal,AndradePr,KrivPr}, an
approximation method in which the spacetime of the merger is
considered to be a perturbation of the spacetime of the single final
hole.
Close limit calculations, to first order in some separation parameter, 
have proven to give excellent agreement with numerical relativity in the 
case of head on collisions, models simple enough to be computed with 
numerical relativity. In principle one can apply the close limit method
to collisions that are still beyond the scope of numerical methods, and in
fact, one such result has already been reported\cite{CLinspiral}.

In principle, first order perturbation theory works in the limit that
some expansion parameter vanishes. In practice, the calculations are
reasonably accurate for a range of that expansion parameter, up to
some maximum. A shortcoming of first order perturbation theory is that
there is no indication, internal to the method, of the size of that
maximum.  For this reason, second order close limit theory was
developed \cite{GNPP1,GNPP2,GNPP3,G_masscrrct,GNPPBY,GNPPboost}.  When
perturbations become large enough that the predictions of first order
calculations differ significantly from the predictions of second
order, it is a sign that perturbation theory is at its limit of
applicability.  Comparison with numerical relativity results, where
they are available, has demonstrated that this method of determining
``error bars'' on perturbation theory is quite
reliable\cite{GNPP1,GNPP2,GNPP3,G_masscrrct,GNPPBY,GNPPboost}.
Perturbation calculations must deal with the freedom to redefine
coordinates, that is to do coordinate ``gauge transformations.'' This
can be approached in two very different ways. One way is to eliminate
the coordinate freedom by fixing the coordinates. This, in fact, was
the way in which the perturbation work on nonrotating
holes\cite{GNPP1,GNPP2,GNPP3,G_masscrrct,GNPPBY,GNPPboost} was carried
out. (Except for very recent work \cite{CampLousto}, these second
order calculations were done as perturbations of a nonrotating black
hole.)  The even parity perturbations that were the focus of these
calculations, were done with a second order extension of the
Zerilli\cite{Z} formalism.  In the extension to second order, the
Regge-Wheeler\cite{RW} conditions were used to fix the coordinates to
second order as well as first. This gauge-fixed approach simplifies
some of enormous complexity of the equations that arise,  but a
price has to be paid and paid twice. Explicit gauge transformations
have to be performed to convert the initial value solution to the
Regge-Wheeler gauge and an explicit transformation has to be performed
to relate the computed perturbations to an asymptotically flat gauge
in which information can be extracted about radiated power.

A second way of dealing with gauge freedom is to find combinations of
metric perturbations that are gauge invariant, and to work only with
gauge invariant quantities. The details of this procedure in first
order calculations were  given by Moncrief\cite{MON}, who
constructed a gauge invariant combination that we shall call the
Moncrief first order invariant.  For vacuum perturbations, this
Moncrief invariant satisfies the same equation as the Zerilli
wavefunction and in the Regge-Wheeler gauge, for vacuum perturbations,
the Moncrief invariant reduces to the Zerilli wavefunction.

Moncrief's choice is not the unique gauge invariant that can be
constructed from even parity metric perturbations, but it is the
unique choice constructed entirely from ``data'' on an initial
hypersurface (i.e., the 3-geometry and the extrinsic curvature). This
property makes it especially convenient for use in calculations of
evolution of perturbations.  In work of this sort we start with a
solution on an initial hypersurface in some gauge (whatever gauge is
convenient for solving the initial value problem).  In the Moncrief
formulation, the starting value of the wave function to be evolved can
be evaluated in any gauge, hence it can immediately and directly be
constructed from the initial value solution. A gauge invariant
formulation is similarly convenient for extracting information about
radiation energy carried in the perturbations. In principle this
requires that one examine the perturbations in an asymptotically flat
gauge.  With a gauge invariant formulation, the form of the invariant
can be related to an asymptotic gauge in a relatively convenient way.

The goal of the present paper is to cast the problem of second order
perturbations of the Schwarzschild spacetime into a form that has the
same convenience that the Moncrief formalism provides in first order
perturbation calculations of evolution. That is, we will provide a
second order formalism in which computations are carried out only with
gauge invariant quantities, and in which these quantities are
constructed from the first and second order perturbations of the
spatial metric and of the extrinsic curvature for an initial
hypersurface. The motivation behind this is to demonstrate the
potential advantages of such a reformulation. To make this
demonstration clear we choose to focus not on the general problem of
perturbations of the Schwarzschild spacetime, but rather on a
restricted class of problems for which a formalism can be presented
with explicit details.

One restriction is that we will consider only the axisymmetric
collision of nonspinning holes. Inherent in this restriction is the
simplification that odd-parity perturbations vanish at every order.
The decomposition of perturbations into multipoles will be used, and
for simplicity we will present results only for the second order
quadrupole (i.e., $\ell=2$) perturbations. This choice is justified by
the fact that most of the radiation is expected to appear in the
$\ell=2$ multipole.

A very different sort of restriction is that we shall consider only
$\ell=2$ {\em first} order perturbations. In principle, first order
multipoles of many different orders can couple to the second order
quadrupole through the nonlinear mixing of first order multipoles. One
justification for ignoring the contributions of $\ell\neq2$ first
order terms is the example of close limit perturbations of collisions
starting with conformally flat time symmetric initial data, such as
the initial data of Brill and Lindquist\cite{BL}, or especially the
solution of Misner\cite{MISNER}, the first and clearest example to
which second order analysis has been
applied\cite{GNPP1,GNPP2,GNPP3}. For these initial data sets in the
case of an axisymmetric collision of two equal mass holes, the first
order perturbations are purely quadrupolar.  For other initial value
solutions we might expect quadrupolar first order perturbations to be
larger in some sense than other first order multipoles, but this is an
inadequate justification for ignoring other first order multipoles.
The real justification then is to simplify the presentation of very
lengthy expressions that illustrate a more generally valid approach.

In the remainder of this section we sketch out the basic
ideas behind the construction of a second order gauge invariant;
details will be given in the sections that follow.

We consider that we have a parameterized family of spacetime metrics
of the form $g_{\alpha\beta}(x^\mu,\epsilon)$, and that these metrics
can be expanded as
\begin{equation}
  \label{eq:deforders}
g_{\alpha\beta}(x^\mu)=g_{\alpha\beta}^{(0)}(x^\mu)
+\epsilon\,g_{\alpha\beta}^{(1)}(x^\mu)+
\frac{1}{2}\epsilon^{2}\,g_{\alpha\beta}^{(2)}(x^\mu)+\cdots\ ,
\end{equation}
where the background metric $g_{\alpha\beta}^{(0)}$ is the
Schwarzschild metric in our case; $g_{\alpha\beta}^{(1)}$ is called
the first order perturbation to the metric; $g_{\alpha\beta}^{(2)}$ is
called the second order perturbation to the metric, and so forth.  Let
us now consider a parameterized family of coordinate transformations,
also called transformations of the coordinate gauge, or simply ``gauge
transformations,'' $x^{\mu}_{new}=F^{\mu}(x^\alpha,\epsilon)$, that can be
expanded as
\begin{equation}
  \label{eq:defgt}
  x^{\mu}_{new}=x^\mu+\epsilon\xi^{(1)\mu}+\frac{1}{2}\epsilon^2\zeta^{(2)\mu}+\cdots
\end{equation}
Such a change of coordinates will transform the metric perturbations
$g_{\alpha\beta}^{(1)},g_{\alpha\beta}^{(2)},\cdots$.
It is useful to consider special cases of the general transformation 
(\ref{eq:defgt}). If $\xi^{(1)\mu}=0$, we call the transformation a purely
second order transformation. Note that 
$g_{\alpha\beta}^{(1)}$ is invariant under this type of transformation, but
$g_{\alpha\beta}^{(2)}, g_{\alpha\beta}^{(3)}\cdots$ are not.

The Moncrief\cite{MON} formalism is based on a certain linear
combination of first order perturbations $g_{\alpha\beta}^{(1)}$ that
can be determined purely from hypersurface information, i.e., from the
first and second fundamental form of a hypersurface that, to zero
order in $\epsilon$ is a constant time surface in the Schwarzschild
geometry. Moncrief shows this combination to be gauge invariant
[invariant under transformation (\ref{eq:defgt})], to carry all the
first order gauge invariant information, and to satisfy a simple wave
equation, the Zerilli\cite{Z} equation.  We use $\Psi^{(1)}$ to denote
Moncrief's combination of first order perturbations
$g_{\alpha\beta}^{(1)}$. [We will present this combination explicitly
below in (\ref{WFFO}) after we have introduced multipole decomposition
and the appropriate notation.]  We will use $L^{(2)}$ to represent the
same combination of second order perturbations
$g_{\alpha\beta}^{(2)}$. It follows immediately that the second order
combination $L^{(2)}$ is invariant under purely second order gauge
transformations. Since $\Psi^{(1)}$ is constructed from the first order
perturbations of the spatial geometry and extrinsic curvature of a
hypersurface, it follows that $L^{(2)}$ can also be constructed from
hypersurface information. We next turn to the question of the wave
equation satisfied by $L^{(2)}$.

The vacuum Einstein equations can be written as
\begin{equation}
\widehat{E}(g_{\alpha\beta})=
\widehat{E}(g_{\alpha\beta}^{(0)}+\epsilon\,g_{\alpha\beta}^{(1)}+
\frac{1}{2}\epsilon^{2}\,g_{\alpha\beta}^{(2)}+\cdots)=0
\label{E}
\end{equation}
where $\widehat{E}$ represents the set of nonlinear differential
operators that generates the Einstein equations.  The terms in
(\ref{E}) that are zero order in $\epsilon$ will be satisfied
automatically because $g_{\alpha\beta}^{(0)}$, the background metric,
is a solution to the vacuum Einstein equations.  To find the equations
satisfied by the first order perturbation, we expand (\ref{E}) in
powers of $\epsilon$, and write the set of first order equations as
\begin{equation}
\widehat{O}(g_{\alpha\beta}^{(1)})=0\ .
\label{FO}
\end{equation}
Since the perturbations $g_{\alpha\beta}^{(1)}$ can only appear
linearly, $\widehat{O}$ represents a set of linear differential operators.

The second order part of the expansion of (\ref{E}) will involve terms
linear in second order perturbations and terms involving products of
first order perturbations. These equations can be written symbolically
as
\begin{eqnarray}
\widehat{O}(g_{\alpha\beta}^{(2)})=
\widehat{S}(g_{\alpha\beta}^{(1)},g_{\alpha\beta}^{(1)})\ .
\label{SO}
\end{eqnarray}
In this form, the products of first order terms appear on the right. 
If the perturbative
problem is solved order by order, the first order problem may be considered 
already to have been solved, so that the right hand side of
(\ref{SO}) can be considered as known.
It should especially be noticed that the operator $\widehat{O}$
is ``zero'' order. That is, $\widehat{O}$ is precisely the same operator
that appears in the first order equations (\ref{FO}).  One can view 
(\ref{SO}) as a system of differential equations differing from 
(\ref{FO}) only by the presence of known source terms.
We know that the first order equations can be rearranged
into a single wave equation, the Zerilli equation, which we symbolize as
\begin{eqnarray}
\widehat{Z}(\Psi^{(1)}) = 0\ .
\label{FOC}
\end{eqnarray} 
It follows that the equations of (\ref{SO})
can be rearranged to give a single wave equation
\begin{equation}
\widehat{Z}(L^{(2)})=
\widehat{S}_{\rm Mon}(g_{\alpha\beta}^{(1)},g_{\alpha\beta}^{(1)})\ ,
\label{SOZ}
\end{equation}
in which the differential operator $\widehat{Z}$ is the Zerilli
operator.  The right hand side represents the set of terms quadratic
in first order perturbations, that result from forming the Zerilli
equation for the Moncrief combination. These first order terms can be
viewed as known once the first order perturbation problem has been
solved, so (\ref{SOZ}) is to be viewed as a wave equation for
$L^{(2)}$ with a known source.

Though the quantity $L^{(2)}$ is constructed from hypersurface
information and satisfies a convenient equation, it is not what we
seek. We have seen that it is invariant under purely second order
perturbations, but it is not invariant under more general gauge
transformations. That is, $L^{(2)}$ will in general change under a
transformation (\ref{eq:defgt}) with $\xi^{(1)}\neq0$.
In order to construct  a second order perturbation function 
that {\em is} gauge
invariant we must add another type of expression to $L^{(2)}$.
Let $Q^{(1)}$ represent any combination of products of first order
perturbations. Since the operator $\widehat{Z}$ is linear, for any such 
$Q^{(1)}$ the quantity
\begin{equation}
  \label{eq:defpsi2}
  \Psi^{(2)}\equiv L^{(2)}+Q^{(1)}
\end{equation}
will satisfy an equation of the form
\begin{equation}
\widehat{Z}(\Psi^{(2)})=
\widehat{Z}(L^{(2)}+Q^{(1)})=
\widehat{S}_{\rm Mon}(g_{\alpha\beta}^{(1)},g_{\alpha\beta}^{(1)})
+ \widehat{Z}(Q^{(1)})\ .
\label{SOC}
\end{equation}
The added source term $\widehat{Z}(Q^{(1)})$ is known once the first
order problem is solved, so $\Psi^{(2)}$, like $L^{(2)}$, satisfies a
Zerilli equation with a known source.  One of the main points of this
paper is to display explicit forms of $Q^{(1)}$ for which $\Psi^{(2)}$
is gauge invariant for the general gauge transformation in
(\ref{eq:defgt}). It should be noted at the outset that $Q^{(1)}$
cannot be unique, and hence $\Psi^{(2)}$ cannot be unique. To see
this, consider $\Xi(\Psi^{(1)}, \Psi^{(1)})$ to be any quadratic
combination of terms in $\Psi^{(1)}$, then define
\begin{equation}\label{altPsi2}
\Psi^{(2)}_{\rm alt}=\Psi^{(2)}+\Xi(\Psi^{(1)}, \Psi^{(1)})\ ,
\end{equation}
and note the following: (i) $\Psi^{(1)}$ is gauge invariant under
first and second order transformations, hence $\Psi_{\rm alt}^{(2)}$
is first and second order gauge invariant if $\Psi^{(2)}$
is first and second order gauge invariant. (ii) Since Moncrief's
$\Psi^{(1)}$ is constructed completely from  hypersurface data, 
it follows that this is true also of $\Psi_{\rm alt}^{(2)}$ if it is
true of $\Psi^{(2)}$. (iii) Like 
the original invariant, the alternate invariant
$\Psi_{\rm alt}^{(2)}$ satisfies a Zerilli wave
equation 
\begin{equation}\label{altzeri}
\widehat{Z}(\Psi^{(2)}_{\rm alt})=
\widehat{Z}\left(L^{(2)}+Q^{(1)}+\Xi(\Psi^{(1)},\Psi^{(1)})
\right)=
\widehat{S}_{\rm Mon}(g_{\alpha\beta}^{(1)},g_{\alpha\beta}^{(1)})
+ \widehat{Z}(Q^{(1)})+\widehat{Z}\left(
\Xi(\Psi^{(1)},\Psi^{(1)})\right)\ ,
\end{equation}
with a known source.

The remainder of this paper will be organized as follows. In
Sec.\,\ref{sec:pertsto2} the perturbed metric tensor and wave
equations, to first and second order, are introduced. The details of
first and second order gauge transformations are discussed in
Sec.\,\ref{sec:1and2inv} and a wave function is presented that is
invariant under these transformations. The procedure for finding
Cauchy data for this wave function is given in Sec.\,\ref{sec:id}. The
relationship of the invariant wavefunction to gravitational wave
energy is analyzed in Sec.\,\ref{sec:energy}, and a summary and
discussion are given in Sec.\,\ref{sec:disc}.  Throughout the paper we
use the conventions of Misner {\it et al.}\cite{MTW}. In particular we
use a metric with sign conventions -+++, and units in which $c=G=1$.


\section{Perturbed metric tensor and Moncrief wave equation for second order}
\label{sec:pertsto2}

\subsection{Perturbed metric tensor}

As was discussed in Sec.\,\ref{sec:intro}, to both first and second
order we consider only $\ell=2$ multipoles. Using the standard
Regge-Wheeler\cite{RW} notation we can write the perturbations [the
$g_{\alpha\beta}^{(1)}$of (\ref{eq:deforders})] to the Schwarzschild
metric for mass $M$, as 
\begin{eqnarray}
g_{tt} &=& -(1-2M/r) \left[1- \left\{\epsilon H_{0}^{(1)}(r,t)+\frac{\epsilon^{2}}{2} H_{0}^{(2)}(r,t)\right\} P_{2}(\theta)\right]\label{METRICstart}\\
g_{tr} &=& \left[\epsilon H_{1}^{(1)}(r,t)+\frac{\epsilon^{2}}{2}H_{1}^{(2)}(r,t)\right] P_{2}(\theta)\\
g_{t\theta} &=& \left[\epsilon h_{0}^{(1)}(r,t)+\frac{\epsilon^{2}}{2}
h_{0}^{(2)}(r,t)\right] P_{2}^{'}(\theta)\\
g_{rr}&=& (1-2M/r)^{-1}\left[1+\left\{\epsilon H_{2}^{(1)}(r,t)+\frac{\epsilon^{2}}{2}
H_{2}^{(2)}(r,t)\right\} P_{2}(\theta)\right]\label{grr}\\
g_{r\theta} &=& \left[\epsilon h_{1}^{(1)}(r,t)+\frac{\epsilon^{2}}{2} h_{1}^{(2)}(r,t)\right] P_{2}^{'}(\theta)\\
g_{\theta\theta} &=& r^2 \left[1+\left\{\epsilon K^{(1)}(r,t)+\frac{\epsilon^{2}}{2}
K^{(2)}(r,t)\right\} P_{2}(\theta) + \left\{\epsilon G^{(1)}(r,t)+\frac{\epsilon^{2}}{2} G^{(2)}(r,t)\right\}P_{2}^{''}(\theta)\right]\label{METRICpenult}\\
g_{\phi\phi} &=& r^2 \left[\sin^2 \theta+\left\{\epsilon K^{(1)}(r,t)+\frac{\epsilon^{2}}{2} K^{(2)}(r,t)\right\}\sin^2 \theta 
P_{2}(\theta)+ \left\{\epsilon G^{(1)}(r,t)+\frac{\epsilon^{2}}{2}
G^{(2)}(r,t)\right\}\sin\theta \cos\theta P_{2}^{'}(\theta)\right].
\label{METRICend}
\end{eqnarray}
Here we are using $P_{2}(\theta)$ to denote the Legendre polynomial of
order 2, with argument $\cos\theta$. By $P_{2}^{'}(\theta)$ and
$P_{2}^{''}(\theta)$ we mean respectively the first and second
derivative of $P_{2}(\theta)$ with respect to $\theta$.
We include an upper index
between parentheses whenever it is necessary to clarify the order of a
perturbation quantity.

\subsection{Zerilli wave equation}

As described in Sec.\,\ref{sec:intro}, a starting point in the search
for second order invariants is the first order Moncrief invariant
$\Psi^{(1)}$. For $\ell=2$, in terms of the Regge-Wheeler notation
introduced above, $\Psi^{(1)}$ is the following linear combination of
first order perturbations
\begin{eqnarray}
\Psi^{(1)} &\equiv&{r \over 6\,(2r+3M)}\left[2 (r-2M) (H_{2}^{(1)}-r \partial_r K^{(1)})-2 (r-3M)
K^{(1)} + 6 \left\{r K^{(1)}+ {(r-2M) \over  r} (r^2 \partial_r G^{(1)}-2 h_{1}^{(1)})\right\}\right]\ .
\label{WFFO}
\end{eqnarray}
Moncrief has shown that this first order combination is invariant under 
gauge transformation (\ref{eq:defgt}) and satisfies the Zerilli\cite{Z}
equation
\begin{equation}\label{FOZ}
\widehat{Z}(\Psi^{(1)})=0\ .
\end{equation}
Here the Zerilli operator $\widehat{Z}$ is 
\begin{equation}\label{ZOP}
\widehat{Z}=-{\partial^2 \over \partial t^2} 
+{\partial^2 \over \partial r_*^2} +V(r) \ ,
\end{equation}
where  $r_*$ is the usual ``tortoise'' coordinate covering the exterior
of the black hole,
\begin{equation}
r_* = r+2 M \ln({r/2M} -1)\ ,
\end{equation}
so that the horizon is at $r_*=-\infty$ and spatial infinity is at $r_*=\infty$.
The potential term in the $\ell=2$ Zerilli operator is  given by
\begin{equation}\label{def:V}
V(r) =6\left(1-{2M \over r}\right) {4r^3+4r^2M+6rM^2+3M^3 \over r^3(2r+3M)^2}\ .
\end{equation}

We now define $L^{(2)}$ to be the second order equivalent of
$L^{(1)}$:
\begin{equation}\label{def:L2}
L^{(2)}\equiv{r \over 6\,(2r+3M)}\left[2 (r-2M) (H_{2}^{(2)}-r \partial_r
K^{(2)})-2 (r-3M) K^{(2)} + 6 \left\{r K^{(2)}+ {(r-2M) \over r} (r^2
\partial_r G^{(2)}-2 h_{1}^{(2)})\right\}\right]\ .
\end{equation}
As was argued in connection with (\ref{SOZ}), this second order
combination satisfies an equation of the form (\ref{SOZ}) 
with $\widehat{S}_{\rm Mon}$ a sum of products of the first order
perturbations,
$H_0^{(1)}$, $H_1^{(1)}$, $H_2^{(1)}$, $h_0^{(1)}$, $h_1^{(1)}$,
$K^{(1)}$, $G^{(1)}$, and the derivatives of these functions.  The
explicit form of $\widehat{S}_{\rm Mon}$ is straightforward to compute; one
simply repeats the steps that lead to the first order Zerilli equation
(\ref{FOZ}) and keeps all terms of second order.
But the result is extremely lengthy and will not be displayed here.

It is worth noting that $\widehat{S}_{\rm Mon}$ is automatically
invariant for purely second order transformations since first order
perturbations do not change for purely second order
transformations. Since $\widehat{Z}$ is invariant under general gauge
transformations, (\ref{SOZ}) then tells us that $L^{(2)}$ must be
gauge invariant under purely second order transformations. This is a
property that also follows from the manner in which $L^{(2)}$ is
constructed. The validity of (\ref{SOZ}) can then be viewed as a check
of consistency.

\section{second order invariant wavefunction}
\label{sec:1and2inv}
To explore the gauge changes in metric perturbations we must introduce
a specific form of a gauge transformation.  The form of the
transformation in (\ref{eq:defgt}) was used in the studies by Gleiser
{\em et al.}\cite{GNPP1,GNPP2,GNPP3}.  Here we choose instead
the equivalent form of higher order gauge transformation given by
Bruni {\em et al.}\cite{MB}
\begin{equation}
x_{new}^{\alpha}=x^{\alpha}+\epsilon\
\xi^{(1)\alpha}(x^{\beta})+ 
(\epsilon^{2}/2)\ \left[\xi^{(2)\alpha}(x^{\beta})+
\partial_{\mu}\xi^{(1)\alpha}(x^{\beta})
\ \xi^{(1)\mu}(x^{\beta})\right]\ .
\end{equation}
The reason for this choice is a practical one. With this notation
$\xi^{(1)\alpha}$ and $\xi^{(2)\alpha}$ can be treated as generating
vectors and the first and second order gauge transformations can be
written in the form of Lie derivatives
\begin{eqnarray}
\delta g_{\alpha\beta}^{(1)}&=&{\cal{L}}_{\xi^{(1)}}\,
g_{\alpha\beta}^{(0)}\\
\delta g_{\alpha\beta}^{(2)}&=&({\cal{L}}_{\xi^{(2)}}+{\cal{L}}_{\xi^{(1)}}^{2})\, g_{\alpha\beta}^{(0)}+
2{\cal{L}}_{\xi^{(1)}}\,g_{\alpha\beta}^{(1)}\ .
\end{eqnarray}
Lie derivatives of tensors can be handled automatically by the Maple
symbolic manipulation language that was used to do the computations.

Since we are using multipole decomposition and keeping only the 
quadrupole terms, we write the components of the generating vectors
as
\begin{eqnarray}
\xi^{(1)}&=&\left\{C_{t}^{(1)}(r,t)\:
P_{2}(\theta),\ C_{r}^{(1)}(r,t)\:P_{2}(\theta),\ C_{\theta}^{(1)}(r,t)
\:P_{2}^{'}(\theta),\ 0\right\}\label{xi1}\\
\xi^{(2)}&=&\left\{C_{t}^{(2)}(r,t)\:P_{2}(\theta),\ C_{r}^{(2)}(r,t)
\:P_{2}(\theta),\ C_{\theta}^{(2)}(r,t)\:P_{2}^{'}(\theta),\ 0\right\}\label{xi2}\ .
\end{eqnarray}
The first order gauge transformations then take the form
\begin{eqnarray}
\delta
H_{2}^{(1)}&=&2\partial_{r}C_{r}^{(1)}(r,t)
-r^{-2}\left(1-2M/r\right)^{-1}2MC_{r}^{(1)}(r,t)
\label{FOGTF}\\
\delta h_{1}^{(1)}&=&\left(1-2M/r\right)^{-1}
C_{r}^{(1)}(r,t)+
r^2\partial_{r}C_{\theta}^{(1)}(r,t)\\
\delta K^{(1)}&=&2r^{-1}C_{r}^{(1)}(r,t)\\
\delta G^{(1)}&=&2C_{\theta}^{(1)}(r,t)\\
\delta H_{1}^{(1)}&=&
\left(1-2M/r\right)^{-1}
\partial_{t}C_{r}^{(1)}(r,t)
-\left(1-2M/r\right)\partial_{r}C_{t}^{(1)}(r,t)\\
\delta h_{0}^{(1)}&=&
-\left(1-2M/r\right)
C_{t}^{(1)}(r,t)+
r^2\partial_{t}C_{\theta}^{(1)}(r,t)\ .
\label{FOGTL}
\end{eqnarray}
Pure second order gauge transformations would look exactly like
(\ref{FOGTF})--(\ref{FOGTL}), replacing upper index $1$ by $2$.
An example of general second order gauge transformations, after
projecting into $\ell=2$, is
\begin{eqnarray}
\delta G^{(2)}&=&2C_{\theta}^{(2)}+\frac{(-2)}{7r^3(r-2M)}
(-r^3(r-2M)C_{t}^{(1)}\partial_{t}C_{\theta}^{(1)}-\nonumber\\
&&3r^2C_{r}^{(1)}C_{r}^{(1)}-18r^3(r-2M)C_{\theta}^{(1)}C_{\theta}^{(1)}+2r^3(r-2M)C_{r}^{(1)}\partial_{r}G^{(1)}-\nonumber\\
&&6r(r-2M)h_{1}^{(1)}C_{r}^{(1)}-6r(r-2M)h_{0}^{(1)}C_{t}^{(1)}+3(r-2M)^2C_{t}^{(1)}C_{t}^{(1)}+2r^3(r-2M)C_{t}^{(1)}\partial_{t}G^{(1)}+\nonumber\\
&&4r^3(r-2M)C_{\theta}^{(1)}K^{(1)}-18r^3(r-2M)C_{\theta}^{(1)}G^{(1)}+\nonumber\\
&&4r^2(r-2M)C_{r}^{(1)}G^{(1)}-r^3(r-2M)C_{r}^{(1)}\partial_{r}C_{\theta}^{(1)}+8r^2(r-2M)C_{\theta}^{(1)}C_{r}^{(1)})\
.
\label{SOGT}
\end{eqnarray}
Expressions like (\ref{SOGT}) 
generate
the first order gauge transformations of the linear part in second
order perturbations of $L^{(2)}$.

 The key to building invariants at second order is to 
take combinations of the gauge transformation equations in 
(\ref{FOGTF})--(\ref{FOGTL}) that isolate the coefficient 
functions occurring in (\ref{xi1}) and (\ref{xi2}),
\begin{eqnarray}
\delta[h_{1}^{(1)}-{r^2 \over 2} \partial_r G^{(1)}]&=&{r \over
(r-2M)}C_{r}^{(1)}\label{IT1}\\
\delta[h_{0}^{(1)}-{r^2 \over 2} \partial_t G^{(1)}]&=&-{(r-2M) \over
r}C_{t}^{(1)}\label{IT2}\\
\delta[G^{(1)}]&=&2C_{\theta}^{(1)}\ . \label{IT3}
\end{eqnarray} 
With these at hand one can construct first order quadratic terms
that, under a first order transformation, cancel gauge dependent terms
arising from the transformation of $L^{(2)}$. This procedure leads 
to the following as the simplest choice for a first and second order
gauge invariant. 
\begin{equation}
\Psi^{(2)}_{\rm RW}\equiv L^{(2)}+Q^{(1)}_{\rm RW}
\end{equation}
where $L^{(2)}$ is the second order equivalent 
of $\Psi^{(1)}$, given in (\ref{def:L2}), and
\begin{eqnarray}\label{def:Q1}
Q^{(1)}_{\rm RW}&\equiv&
{4 \over 21 r^3 (r-2M)(2r+3M) } \left[ (r-2M)^2 {(h_{1}^{(1)}-{r^2
\over 2} \partial_r G^{(1)})} (-2r(3r+M)K^{(1)}+6r(r-2M)\partial_r
h_{1}^{(1)}+\nonumber\right.\\&&
6r(5r+4M)G^{(1)}-7r^2M\partial_r 
 K^{(1)}+r(3r+2M)H_{2}^{(1)}-3r^3(r-2M)\partial_{r^2}
G^{(1)}-6(5r+4M)h_{1}^{(1)}+\nonumber\\&&
6r^2(r+5M)\partial_r G^{(1)}+r^3(r-2M)\partial_{r^2}
K^{(1)}-r^2(r-2M)\partial_r H_{2}^{(1)})+\nonumber\\&&
r^2(r-2M)^2 {\partial_r({(r-2M) \over r}  (h_{1}^{(1)}-{r^2 \over 2}
\partial_r G^{(1)}))} (-2r H_{2}^{(1)}+6 h_{1}^{(1)}-3r^2 \partial_r
G^{(1)}-6r G^{(1)}+
r^2 \partial_r K^{(1)}+2r K^{(1)})+\nonumber\\&&
{r^2 (r-2M)\over 2} { G^{(1)}} (18(r-2M) h_{1}^{(1)}-3r^2(r-2M)\partial_r
G^{(1)}-3r^2(r-2M) \partial_r K^{(1)}-24r(2r+3M) G^{(1)}+\nonumber\\&&
9r(2r+3M)K^{(1)}-3r(r-2M) H_{2}^{(1)})+
{6r^2(r-2M)^2 \over 2} { (r^2 G^{(1)}-(r-2M)
h_{1}^{(1)})}\partial_r G^{(1)}+\nonumber\\&&
 r(r-2M)({(-r) \over (r-2M)} { (h_{0}^{(1)}-{r^2 \over 2} \partial_t
G^{(1)})}) (-r^2(r-2M) \partial_t H_{2}^{(1)}+r^3(r-2M) \partial_{rt}
K^{(1)}-r^2(2r+3M) \partial_t K^{(1)}-\nonumber\\&&
3r^3(r-2M) \partial_{rt}
G^{(1)}+3r(r-2M) \partial_t h_{1}^{(1)}+3r(r-2M) H_{1}^{(1)}+3r(r-2M)
\partial_r h_{0}^{(1)}+6r^2(2r+3M) \partial_t G^{(1)}-\nonumber\\&&
6(5r+4M)h_{0}^{(1)})+\nonumber\\&& 
r^2(r-2M)^2 { \partial_r ({(-r) \over (r-2M)} (h_{0}^{(1)}-{r^2
\over 2} \partial_t G^{(1)}))} (-3r^2 \partial_t G^{(1)}+6 h_{0}^{(1)}-2(r-2M)
H_{1}^{(1)}+r^2 \partial_t K^{(1)})+\nonumber\\&&
{(r-2M)^2 \over (2r)} (38r^2+28rM+4M^2){ (h_{1}^{(1)}-{r^2 \over 2}
\partial_r G^{(1)})} (h_{1}^{(1)}-{r^2 \over 2} \partial_r G^{(1)})-\nonumber\\&&
8(r-2M)^2r^2  { (h_{1}^{(1)}-{r^2 \over 2} \partial_r G^{(1)})}
\partial_r ({(r-2M) \over r}(h_{1}^{(1)}-{r^2 \over 2} \partial_r G^{(1)}))-\nonumber\\&&
6r(r-2M)^2(5r+4M) (h_{1}^{(1)}-{r^2 \over 2} \partial_r G^{(1)})
{ G^{(1)}}+\nonumber\\&&
3r^2(r-2M)^3 { (h_{1}^{(1)}-{r^2 \over 2} \partial_r G^{(1)})}
\partial_r G^{(1)}+\nonumber\\&&
r^3(r-2M)^2 { \partial_r ( {(r-2M) \over r}(h_{1}^{(1)}-{r^2 \over 2} \partial_r G^{(1)}))}
\partial_r ({(r-2M) \over r}(h_{1}^{(1)}-{r^2 \over 2} \partial_r G^{(1)}))+\nonumber\\&&
6r^3(r-2M)^2 \partial_r ({(r-2M) \over r}(h_{1}^{(1)}-{r^2 \over 2} \partial_r G^{(1)}))
{ G^{(1)}}+\nonumber\\&&
6r^3(2r+3M)(r-2M) { G^{(1)}} G^{(1)}-3r^4(r-2M)^2 \partial_r G^{(1)}
{ G^{(1)}}+{3 \over 4} r^4(r-2M)^3 { \partial_r G^{(1)}} \partial_r
G^{(1)}-\nonumber\\&&
3(5r+3M)(r-2M)^2 ({(-r) \over (r-2M)} { (h_{0}^{(1)}-{r^2 \over 2} \partial_t
G^{(1)})}) ({(-r) \over (r-2M)} (h_{0}^{(1)}-{r^2 \over 2} \partial_t
G^{(1)}))+\nonumber\\&&
6r(r-2M)^3 ({(-r) \over (r-2M)} { (h_{0}^{(1)}-{r^2 \over 2} \partial_t
G^{(1)})}) \partial_r ({(-r) \over (r-2M)} (h_{0}^{(1)}-{r^2 \over 2} \partial_t
G^{(1)}))-\nonumber\\&&\left.r(r-2M)^4 { \partial_r ({-r \over (r-2M)}
(h_{0}^{(1)}-{r^2 \over 2} \partial_t G^{(1)}))} \partial_r ({-r \over (r-2M)}
(h_{0}^{(1)}-{r^2 \over 2} \partial_t G^{(1)}))\right]\ .
\end{eqnarray}
In the Regge-Wheeler choice of gauge\cite{RW} the perturbation
functions $h_{0}^{(1)}$, $h_{1}^{(1)}$ and $G^{(1)}$ are chosen to
vanish. In this gauge the right hand side of (\ref{def:Q1})
vanishes. For this reason we denote by
$Q^{(1)}_{\rm RW}$
the particular choice of quadratic
terms  appearing in (\ref{def:Q1}) and we denote the
corresponding wave function as $\Psi^{(2)}_{\rm RW}$. Note that
$\Psi^{(2)}_{\rm RW}$ becomes simply $L^{(2)}$ in the Regge-Wheeler
gauge. The wave function $\Psi^{(2)}_{\rm RW}$ satisfies a wave
equation
\begin{equation}\label{eq:PsiRW}
\widehat{Z}(\Psi^{(2)}_{\rm RW})=
\widehat{Z}(L^{(2)}+Q^{(1)}_{\rm RW})=
\widehat{S}_{\rm Mon}(g_{\alpha\beta}^{(1)},g_{\alpha\beta}^{(1)})
+ \widehat{Z}(Q^{(1)}_{\rm RW})\equiv{\cal S}_{\rm RW}\ .
\end{equation}
If one works explicitly in the Regge-Wheeler gauge, then the terms
in $\widehat{Z}(Q^{(1)}_{\rm RW})$ vanish.

The explicit expression for the source term is extremely lengthy. To compress
it into a manageable form we introduce a few simplifications in notation. We
use a prime ($'$) to denote partial differentiation with respect to $r$ and 
a dot ($\dot{\ }$) to denote partial differentiation with respect to time,
and we write it in terms of two first order quantities $K$ and $H_2$:
\begin{eqnarray}\label{explicitS}
S_{\rm RW}&=& (-2/189)\left[
(1512M^5-4068r^3M^2+1602r^4M-252r^5-4248rM^4+5490r^2M^3) H_{2} K^{'}
+\right. \nonumber\\&& (234r^4-972r^3M+324M^4-468rM^3+1161r^2M^2) K K+
\nonumber\\&& (30r^6-189r^2M^4+147r^3M^3-77r^4M^2+432rM^5-52r^5M)
\dot{K}\dot{K} + \nonumber\\&&
(-9r^6+72r^5M-216r^4M^2+288r^3M^3-144r^2M^4) K^{'} K^{'}+
\nonumber\\&& (-36r^6-864r^4M^2+288r^5M+1152r^3M^3-576r^2M^4) K
K^{''}+ \nonumber\\&&
(729r^4M^2-459r^5M+72r^6+882r^3M^3+2376rM^5-3348r^2M^4) H_{2}^{'}
K^{'} + \nonumber\\&&
(-783r^4M^2+90r^5M+2538r^3M^3+1944rM^5-3636r^2M^4) H_{2} K^{''}+
\nonumber\\&& (-306r^4M+1944r^3M^2-864M^5+3744rM^4-4320r^2M^3) H_{2}
H_{2}^{'} + \nonumber\\&&
(1575r^2M^2+387r^4-1359Mr^3+3132M^4-2376rM^3) H_{2} H_{2}+
\nonumber\\&& (-450r^3M+1080M^4+2070r^2M^2-2880rM^3) H_{2} K +
\nonumber\\&& (-36r^6-87r^4M^2+48r^5M+540r^2M^4) H_{2} \ddot{K}+
\nonumber\\&& (36r^6-276r^4M^2+162r^3M^3+540r^2M^4-42r^5M) K \ddot{K}+
\nonumber\\&& (-117r^6M+18r^7-576r^3M^4+216r^5M^2+72r^4M^3+432r^2M^5)
K^{'} H_{2}^{''} + \nonumber\\&&
(-117r^6M+18r^7+72r^4M^3-576r^3M^4+432r^2M^5+216r^5M^2) H_{2} K^{'''}
+ \nonumber\\&& (-56r^6M^3-2r^9+48r^7M^2-3r^8M-96r^5M^4+144r^4M^5)
\dot{K} \dot{K^{'''}} + \nonumber\\&&
(-54r^6M+12r^7+120r^4M^3-144r^3M^4+36r^5M^2) K \ddot{H_{2}^{'}}+
\nonumber\\&& (-36r^6M^2+54r^7M-12r^8+144r^4M^4-120r^5M^3) K
\ddot{K^{''}}+ \nonumber\\&& (-60r^6M^3-18r^7M^2-6r^9+27r^8M+72r^5M^4)
K^{'} \ddot{K^{''}}+ \nonumber\\&&
(-18r^7M^2-6r^9+27r^8M+72r^5M^4-60r^6M^3) K^{''} \ddot{K^{'}}+
\nonumber\\&& (-48r^7M^2-16r^6M^3+26r^8M-4r^9-96r^4M^5+128r^5M^4)
\dot{K^{'}} \dot{H_{2}^{''}}+ \nonumber\\&&
(16r^7M^3+4r^{10}-128r^6M^4-26r^9M+48r^8M^2+96r^5M^5) \dot{K^{'}}
\dot{K^{'''}} + \nonumber\\&&
(3r^7M-48r^6M^2+2r^8+96r^4M^4-144r^3M^5+56r^5M^3) \dot{K}
\dot{H_{2}^{''}} + \nonumber\\&&
(151r^7M-258r^6M^2-22r^8+1288r^4M^4-960r^3M^5-316r^5M^3) \dot{K^{'}}
\dot{H_{2}^{'}} + \nonumber\\&&
(18r^7+588r^3M^4-720r^2M^5-133r^5M^2+124r^4M^3-29r^6M) \dot{H_{2}^{'}}
\dot{K} + \nonumber\\&&
(-108r^5+792r^4M-2214r^3M^2+432M^5-1800rM^4+2916r^2M^3) K K^{'}+
\nonumber\\&& (171r^6M-24r^7+2376r^2M^5-1362r^3M^4-211r^4M^3-97r^5M^2)
\dot{K^{'}} \dot{K} + \nonumber\\&&
(57r^6M-6r^7+378r^3M^4-99r^4M^3-135r^5M^2) K^{'} \ddot{K}+
\nonumber\\&& (432M^5-36r^4M+162r^3M^2-360rM^4-108r^2M^3) H_{2}^{'} K+
\nonumber\\&& (-18r^6-72r^3M^3+117r^5M-216r^4M^2-432rM^5+576r^2M^4)
H_{2} H_{2}^{''} + \nonumber\\&&
(36r^6+144r^3M^3-234r^5M+432r^4M^2-1152r^2M^4+864rM^5) K H_{2}^{''} +
\nonumber\\&& (-128r^6M^4+4r^{10}+16r^7M^3+48r^8M^2-26r^9M+96r^5M^5)
\dot{K^{''}} \dot{K^{''}} + \nonumber\\&&
(160r^6M^2-17r^8-6r^7M+1992r^3M^5-2264r^4M^4+474r^5M^3) \dot{K^{'}}
\dot{K^{'}}+ \nonumber\\&&
(4r^8-26r^7M+96r^3M^5-128r^4M^4+48r^6M^2+16r^5M^3) \dot{H_{2}^{'}}
\dot{H_{2}^{'}} + \nonumber\\&&
(48r^6M^2-26r^7M+4r^8+96r^3M^5-128r^4M^4+16r^5M^3) \dot{H_{2}}
\dot{H_{2}^{''}}+ \nonumber\\&&
(-16r^6M^3-48r^7M^2-4r^9+26r^8M-96r^4M^5+128r^5M^4) \dot{H_{2}}
\dot{K^{'''}} + \nonumber\\&&
(-18r^6M^2+27r^7M-6r^8+72r^4M^4-60r^5M^3) H_{2}^{'} \ddot{K^{'}}+
\nonumber\\&& (-54r^6M+12r^7-144r^3M^4+120r^4M^3+36r^5M^2) H_{2}
\ddot{H_{2}^{'}} + \nonumber\\&&
(-57r^6M+6r^7-252r^3M^4+120r^4M^3+93r^5M^2) K^{'} \ddot{H_{2}}+
\nonumber\\&& (-36r^6M^2+144r^4M^4) H_{2} \ddot{K^{''}}+ \nonumber\\&&
(121r^7M-18r^8-816r^3M^5+1120r^4M^4-192r^6M^2-312r^5M^3) \dot{H_{2}}
\dot{K^{''}}+ \nonumber\\&& (33r^7M-6r^8+108r^4M^4-72r^5M^3-33r^6M^2)
K^{''} \ddot{K}+ \nonumber\\&&
(180r^6M-24r^7-384r^4M^3+720r^3M^4-252r^5M^2) K \ddot{K^{'}}+
\nonumber\\&& (-27r^6M+6r^7-72r^3M^4+60r^4M^3+18r^5M^2) H_{2}^{'}
\ddot{H_{2}}+ \nonumber\\&&
(-144r^6M^2+117r^7M-18r^8-252r^5M^3+432r^4M^4) K^{'} \ddot{K^{'}} +
\nonumber\\&& (-32r^6M^3-96r^7M^2+52r^8M-8r^9-192r^4M^5+256r^5M^4)
\dot{H_{2}^{'}} \dot{K^{''}}+ \nonumber\\&&
(232r^6M^3+384r^7M^2-179r^8M+22r^9+1296r^4M^5-1568r^5M^4) \dot{K^{'}}
\dot{K^{''}}+ \nonumber\\&&
(182r^6M^2+43r^7M-18r^8+1224r^3M^5-672r^4M^4-362r^5M^3) \dot{K}
\dot{K^{''}}+ \nonumber\\&& (54r^7M-12r^8-120r^5M^3) H_{2}
\ddot{K^{''}}+ \nonumber\\&& (-27r^7M+6r^8-72r^4M^4+60r^5M^3+18r^6M^2)
K^{''} \ddot{H_{2}} + \nonumber\\&&
(60r^7+2604r^3M^4-1584r^2M^5+67r^5M^2-988r^4M^3-133r^6M) \dot{H_{2}}
\dot{K^{'}}+ \nonumber\\&& (12r^7+108r^3M^4+36r^4M^3-69r^5M^2-12r^6M)
H_{2}^{'} \ddot{K} + \nonumber\\&&
(-42r^7+720r^3M^4-492r^4M^3-216r^5M^2+225r^6M) H_{2} \ddot{K^{'}} +
\nonumber\\&& (-24r^6-78r^3M^3-120r^5M+168r^4M^2-576rM^5+1116r^2M^4)
\dot{H_{2}} \dot{K} + \nonumber\\&&
(36r^7-1152r^3M^4+144r^4M^3+432r^5M^2-234r^6M+864r^2M^5) H_{2}^{'}
K^{''} + \nonumber\\&&
(-36r^6-144r^3M^3-432r^4M^2+1152r^2M^4-864rM^5+234r^5M) H_{2}^{'}
H_{2}^{'} + \nonumber\\&& (18r^6M^2-27r^7M+6r^8-72r^4M^4+60r^5M^3)
K^{'} \ddot{H_{2}^{'}} + \nonumber\\&&
(18r^7+480r^2M^5-840r^3M^4+396r^4M^3-93r^6M+66r^5M^2) \dot{H_{2}}
\dot{H_{2}^{'}}+ \nonumber\\&&
(444r^3M^3+42r^4M^2-195r^5M+54r^6-360r^2M^4) H_{2} \ddot{H_{2}} +
\nonumber\\&& (-660r^2M^4+364r^3M^3-97r^4M^2+94r^5M-33r^6+192rM^5)
\dot{H_{2}} \dot{H_{2}} + \nonumber\\&&
\left.(150r^4M^2+120r^3M^3-60r^5M-360r^2M^4) K \ddot{H_{2}} \right] /
r^2 (2r+3M)^2 (r-2M)\ .
\end{eqnarray}
The quantities $K$ and $H_2$ occurring in 
(\ref{explicitS}) are the following combinations of first order metric
perturbations:
\begin{eqnarray}
K & \equiv & {K}^{(1)} +(r-2M) \left(
\partial_r{G}^{(1)}  -{2 \over r^2}{h}^{(1)}_1 \right) \\
H_2 & \equiv &
{H}^{(1)}_2 + (2r-3M) \left(\partial_r{G}^{(1)} - {2 \over
r^2} {h}^{(1)}_1 \right) + r(r-2M)
\partial_r\left(\partial_r{G}^{(1)} - {2 \over r^2} {h}^{(1)}_1
\right)\ .
\end{eqnarray}
Note that $K$ and $H_2$ reduce to the metric perturbations ${K}^{(1)}$
and ${H}^{(1)}_2$ in the RW gauge.

As explained in connection with (\ref{altPsi2}), it is easy to
construct an alternative wave function by adding to $\Psi^{(2)}_{\rm
RW}$ terms quadratic in $\Psi^{(1)}$. Any such modification, however, will
not have the property that quadratic terms vanish in the Regge-Wheeler gauge, and might make the expression for the source more complicated.

\section{Initial data}
\label{sec:id}

\subsection{Unit normal  to the perturbed hypersurfaces}

Our spacetime will be foliated, outside the horizon, by spacelike
surfaces of constant coordinate time $t$ that, to zero order in
$\epsilon$, agree with the surfaces of constant Schwarzschild
coordinate time. Our initial hypersurface will be assumed to be one of
these constant $t$ surfaces.  Viewed as a foliation of a given
spacetime, the constant $t$ surfaces
will change under coordinate transformation, but our variables
$\Psi^{(1)}$ and $\Psi^{(2)}_{\rm RW}$, specified at a particular coordinate
value, are invariant under such transformation.  More subtle is
the fact that for a fixed foliation of a fixed spacetime, the meaning
of the partial derivative $\partial/\partial t$ will change when gauge
transformations are made that perturbatively change the spatial
coordinate labels on a hypersurface. But our quantities are invariant
with respect to such perturbative diffeomorphisms on the
hypersurfaces, so the $\partial/\partial t$ operation on our
quantities will be unchanged.

To relate the time derivatives of our quantities to the extrinsic
curvature we must first find, to first and second order, the
components of the future directed unit normal to the constant $t$
hypersurfaces.  The computation starts with the relationship of the
metric and the shift components $N_{i}$ in the standard ADM\cite{ADM}
decomposition
\begin{eqnarray}
g_{tr} &=& N_{r}(r,t) \label{SHIFTdefr}\\
g_{t\theta} &=& N_{\theta}(r,t)\label{SHIFTdefth}\\
g_{t\phi} &=& N_{\phi}(r,t)\ . 
\label{SHIFTdefphi}
\end{eqnarray}
 With the definition of the perturbed metric tensor
(\ref{METRICstart})--(\ref{METRICend}) and the definitions for the
shift (\ref{SHIFTdefr})--(\ref{SHIFTdefphi}), the perturbative
expressions for the shift components given to second order are
\begin{eqnarray}
N_{r}(r,t) &=& \left[\epsilon H_{1}^{(1)}+(\epsilon^{2}/2)
H_{1}^{(2)}\right] P_{2}(\theta)\label{SHIFTr}\\ N_{\theta}(r,t) &=&
\left[\epsilon h_{0}^{(1)}+(\epsilon^{2}/2) h_{0}^{(2)}\right]
P_{2}^{'}(\theta)\label{SHIFTth}\\ N_{\phi}(r,t)&=&0\ .
\label{SHIFTphi}
\end{eqnarray}
The vanishing of the shift component $N_{\phi}$ is due to the assumed
axisymmetric character of the collision.

  In order to
obtain the expression for the perturbed lapse to second order we can
use the ADM \cite{ADM} equality  for $^{(4)}g_{tt}$,
\begin{equation}
^{(4)}g_{tt}=-N^{2}+ \, ^{(3)}g^{ij} N_{i} N_{j}\ .
\label{LAPSEDEF}
\end{equation}
Here summation on $i$ and $j$ ranges over $r$, $\theta$ and $\phi$,
and 
$^{(3)}g^{ij}$ is the inverse of the 3-metric on a constant $t$
hypersurface.
Using (\ref{SHIFTr})--(\ref{SHIFTphi}) in (\ref{LAPSEDEF}) gives the
perturbed lapse to second order
\begin{displaymath}
N(r,t) = \sqrt{1-2M/r} \:\: \left[1-\frac{1}{2}\epsilon
H_{0}^{(1)}-\frac{1}{2}\epsilon^{2} (H_{0}^{(2)}
P_{2}(\theta)-H_{1}^{(1)}\,H_{1}^{(1)}P_{2}(\theta)P_{2}(\theta)-
\right.
\end{displaymath}
\begin{equation}\label{LAPSE}
\left. 
{h_{0}^{(1)}h_{0}^{(1)}P_{2}^{'}(\theta)P_{2}^{'}(\theta)\over
r(r-2M)}+
\frac{1}{4}H_{0}^{(1)}H_{0}^{(1)}P_{2}(\theta)P_{2}(\theta)\right]\ .
\end{equation}
In a similar manner one can compute $N^{r}$ and $N^{\theta}$ to second
order from $N^{i}=\:^{(3)}g^{ij}N_{j}$. From these, and $N$, the
covariant $n_{\alpha}$ and contravariant $n^{\alpha}$ components of
the orthonormal vector to the time slices are given as:
\begin{equation}
n_{\alpha} = (-N,0,0,0)
\end{equation}
\begin{equation}
n^{\alpha} = ({1 \over N},-{N^{r} \over N},-{N^{\theta} \over N},0)\ .
\label{OV}
\end{equation}

\subsection{Multipole decomposition of extrinsic curvature} In computing the extrinsic curvature
$K_{ab}$ on our constant $t$ hypersurfaces, it is useful to start with
a tensor harmonic decomposition, similar to that for the metric tensor
given in (\ref{METRICstart})-(\ref{METRICend}). We define the
quantities $K^{(i)}_{rr}$, $K^{(i)}_{r\theta}$, $K^{(i)}_{K}$, and
$K^{(i)}_{G}$ by the relations
\begin{eqnarray}
K_{rr}&=& (1-2M/r)^{-1}\left[\epsilon K_{rr}^{(1)}(r)+\frac{\epsilon^{2}}{2}
K_{rr}^{(2)}(r)\right] P_{2}(\theta)\label{ECFF}\\
K_{r\theta} &=& \left[\epsilon K_{r\theta}^{(1)}(r)+\frac{\epsilon^{2}}{2}
K_{r\theta}^{(2)}(r)\right] P_{2}^{'}(\theta)\\
K_{\theta\theta} &=& r^2 \left[\left\{\epsilon K_{K}^{(1)}(r)+\frac{\epsilon^{2}}{2}
K_{K}^{(2)}(r)\right\} P_{2}(\theta) + \left\{\epsilon
K_{G}^{(1)}(r)+\frac{\epsilon^{2}}{2}
K_{G}^{(2)}(r)\right\}P_{2}^{''}(\theta)\right]\label{ECTHTH}\\
K_{\phi\phi} &=& r^2 \left[\left\{\epsilon K_{K}^{(1)}(r)+\frac{\epsilon^{2}}{2} K_{K}^{(2)}(r)\right\}\sin^2 \theta 
P_{2}^{'}(\theta)+ \left\{\epsilon K_{G}^{(1)}(r)+\frac{\epsilon^{2}}{2}
K_{G}^{(2)}(r)\right\}\sin\theta \cos\theta P_{2}^{'}(\theta)\right]\ .
\label{ECLF}
\end{eqnarray}

\subsection{Time derivatives of the metric perturbations}

To relate the extrinsic curvature and time derivative of metric
perturbations we will need the projector
\begin{equation}
P_{i}^{\mu} = \delta_{i}^{\mu}+ n_{i} n^{\mu}\ .
\end{equation}
into the 3-dimensional hypersurfaces. The extrinsic curvature
is defined, in terms of the unit normal $n^{\alpha}$, as
\begin{equation}
K_{ij} = -\frac{1}{2}P_{i}^{\mu} P_{j}^{\nu} n_{(\mu;\nu)}\ .
\label{KEQ}
\end{equation}
Here the symbol ``$ ;$ '' means covariant derivative with respect to
the 4-metric tensor.  With the relations in
(\ref{SHIFTdefr})-(\ref{OV}) the right hand side of (\ref{KEQ}) can be
decomposed into multipoles with coefficients expressed in terms of
metric perturbations. With (\ref{KEQ}) these coefficients are then
related to the multipole coefficients defined in (\ref{ECFF}) --
(\ref{ECLF}).  As examples of the results of this procedure, the time
derivatives of $H_{2}^{(1)}$ and $H_{2}^{(2)}$ are
\begin{equation}
\partial_{t}H_{2}^{(1)} = 2 (1-2M/r)\: \partial_{r}H_{1}^{(1)}+
(2M/r^2)\: H_{1}^{(1)}+ 2 \sqrt{1-2M/r}\: K_{rr}^{(1)}\ ,
\label{H2FTD}
\end{equation}
\begin{eqnarray}
\partial_{t}H_{2}^{(2)}&=&2 (1-2M/r)\: \partial_{r}H_{1}^{(2)}+
(2M/r^2)\: H_{1}^{(2)}+ 2 \sqrt{1-2M/r}\: K_{rr}^{(2)}
-(2/7r^4)\: \left[12 M h_{1}^{(1)}h_{0}^{(1)}-6r^2
\,h_{0}^{(1)}H_{2}^{(1)}+\right.\nonumber\\&& 
12r^2h_{0}^{(1)}\partial_{r}h_{1}^{(1)}-
24r\,h_{0}^{(1)}\partial_{r}h_{1}^{(1)}-
2r^4\,H_{0}^{(1)}\partial_{r}H_{1}^{(1)}+
2r^4\,H_{1}^{(1)}\partial_{r}H_{2}^{(1)}-
4r^3M\,H_{1}^{(1)}\partial_{r}H_{2}^{(1)}+\nonumber\\&&
\left.4r^3M\,H_{0}^{(1)}\partial_{r}H_{1}^{(1)}+
r^4H_{0}^{(1)}\partial_{t}H_{2}^{(1)}-2r^2M\,H_{1}^{(1)}H_{0}^{(1)}\right]\ .
\label{H2STD}
\end{eqnarray}
These examples show the general pattern. Time derivatives of first
order perturbations of the metric tensor are expressed as linear
combinatins of first order metric perturbations. The expressions for
time derivatives of second order perturbations contain terms linear in
second order metric perturbations, and terms quadratic in first order
perturbations. 

\subsection{Extrinsic curvature and momentum constraints} Since
axisymmetry is assumed, the nontrivial first order momentum constraint
equations are $R_{tr}^{(1)}=0$ and $R_{t\theta}^{(1)}=0$. The partial
derivatives, with respect to time, of the first order metric
perturbations can be reexpressed in terms of components of extrinsic
curvature by using relations like (\ref{H2FTD}). The resulting,
simplified, momentum constraints are\begin{eqnarray}
0&=&r^2(r-2M)\,K_{rr}^{(1)}-r^2(r-2M)\,K_{G}^{(1)}+r^2(r-2M)\,K_{K}^{(1)}+(-2r^2+7rM-6M^2)\,K_{r\theta}^{(1)}-r^2(r-2M)^2\,\partial_{r}K_{r\theta}^{(1)}\label{MC1}\\
0&=&-r^2(r-2M)^2\,\partial_{r}K_{K}^{(1)}+3r^2(r-2M)^2\,\partial_{r}K_{G}^{(1)}+
r(r-2M)^2\,K_{rr}^{(1)}-r(r-2M)^2\,K_{K}^{(1)}+3r(r-2M)^2\,K_{G}^{(1)}\nonumber\\&&
-3(r-2M)^2\,K_{r\theta}^{(1)}\ .\label{MC2}
\end{eqnarray}
These results will be needed below to help connect 
$\partial_{t}\Psi^{(2)}_{\rm RW}$ to the initial value data.

\subsection{Wave function on the initial hypersurface}
 Cauchy data for the first and second order wave equations requires 
the values of $\Psi^{(1)}$ and $\Psi^{(2)}_{\rm RW}$ on the initial hypersurface.
If the 3-geometry of the initial hypersurface is known, $\Psi^{(1)}$
follows immediately from the definition in 
(\ref{WFFO}), and $L^{(2)}$ similarly follows immediately from
(\ref{def:L2}). But the specification of $\Psi^{(2)}_{\rm RW}$ requires
$Q^{(1)}_{\rm RW}$, and the definition $Q^{(1)}_{\rm RW}$ in
(\ref{def:Q1})
appears to require a knowledge of terms that do not follow directly
from the hypersurface 3-geometry. All such terms can be grouped into
only two combinations of perturbations:
\begin{eqnarray}
{ [h_{0}^{(1)}-{r^2\over 2} \partial_t G^{(1)}]}
\mbox{\rm\quad\quad    and\quad\quad}
{[H_{1}^{(1)}-\frac{r^2}{2(r-2M)}\partial_{t}K^{(1)}]}\ .
\label{OFFterms}
\end{eqnarray}
These two first order expressions, however, appear in the
following components of the extrinsic curvature:
\begin{eqnarray}
K_{\theta\theta}^{(1)}&=&\frac{1}{\sqrt{1-{2M \over r}}}
[\ { (h_{0}^{(1)}-{r^2\over 2} \partial_tG^{(1)})} P_{2}^{''}(\theta)+
(r-2M)\ { (H_{1}^{(1)}-\frac{r^2}{2(r-2M)}\partial_{t}K^{(1)})} P_{2}(\theta)]\\
K_{\phi\phi}^{(1)}&=&\frac{1}{\sqrt{1-{2M \over r}}}
[\sin\theta\cos\theta\ { (h_{0}^{(1)}-{r^2\over 2} \partial_tG^{(1)})} P_{2}^{'}(\theta)+
(r-2M) \sin^2\theta\ {(H_{1}^{(1)}-\frac{r^2}{2(r-2M)}\partial_{t}K^{(1)})}P_{2}(\theta)]\ .
\end{eqnarray}
An initial value solution in Einstein's theory consists of both the 3-geometry and extrinsic curvature of the initial hypersurface. From an initial
value solution, then, the terms in (\ref{OFFterms}) can be evaluated, and
the process of specifying $\Psi^{(1)}$ and $\Psi^{(2)}_{\rm RW}$ can be completed.

\subsection{Time derivative of the wave function on the 
initial hypersurface}

The complete specification of 
Cauchy data for the first and second order Zerilli equations includes
the time derivatives $\partial_{t}\Psi^{(1)}$ and
$\partial_{t}\Psi^{(2)}_{\rm RW}$.
These require the time derivatives of metric
perturbations, which 
are found starting with (\ref{KEQ}), as shown in
examples (\ref{H2FTD}) and (\ref{H2STD}). With this approach the
computation of $\partial_{t}\Psi^{(1)}$ is straightforward.  The
computation of $\partial_{t}\Psi^{(2)}_{\rm RW}$, however, requires
$\partial_{t}Q^{(1)}_{\rm RW}$ and is not straightforward.  The
evaluation of this time derivative produces terms that involve
$H_{0}^{(1)}$ multiplied by time derivatives of groups of extrinsic
curvature terms. These groups of terms turn out to be those that occur
in the momentum constraint (\ref{MC1}), so that the troublesome term
is guaranteed to vanish. 
The resulting, simplified expressions for $\partial_{t}\Psi^{(1)}$
and $\partial_{t}\Psi^{(2)}_{\rm RW}$, in terms of hypersurface information,
are
\begin{eqnarray}
\partial_{t}\Psi^{(1)}&=&\frac{2}{21r^4(2r+3M)\sqrt{1-{2M/r}}}
\left[(28r^6M-28r^5M^2-7r^7)\partial_{r}K_{K}^{(1)}+
(-84r^6M+21r^7+84r^5M^2)\partial_{r}K_{G}^{(1)}+\right.\nonumber\\&&
(28r^4M^2-28r^5M+7r^6)K_{rr}^{(1)}+
(168r^4M-168r^3M^2-42r^5)K_{r\theta}^{(1)}+\nonumber\\&&
\left.(14r^6-28r^4M^2-14r^5M)K_{K}^{(1)}+(21r^5M-42r^4M^2)K_{G}^{(1)}\right]
\label{dPsi1dt}
\end{eqnarray}

\begin{eqnarray}
\partial_{t}\Psi^{(2)}_{\rm RW}&=&\frac{2}{21r^4(2r+3M)\sqrt{1-{2M/r}}}
\left[(28r^6M-28r^5M^2-7r^7)\partial_{r}K_{K}^{(2)}+
(-84r^6M+21r^7+84r^5M^2)\partial_{r}K_{G}^{(2)}+\right.\nonumber\\&&
(28r^4M^2-28r^5M+7r^6)K_{rr}^{(2)}+
(168r^4M-168r^3M^2-42r^5)K_{r\theta}^{(2)}+\nonumber\\&&
(14r^6-28r^4M^2-14r^5M)K_{K}^{(2)}+(21r^5M-42r^4M^2)K_{G}^{(2)}+\nonumber\\&&
(4r^8-32r^7M-128r^5M^3+96r^6M^2+64r^4M^4) \partial_{r^2}h_{1}^{(1)}
\partial_{r}K_{G}^{(1)} +\nonumber\\&&  
(4r^7-32r^4M^3-24r^6M+48r^5M^2)\partial_{r^2}K^{(1)}K_{r\theta}^{(1)}+
\nonumber\\&&(-4r^9-80r^5M^4-108r^7M^2+34r^8M+152r^6M^3)
\partial_{r^3}G^{(1)}K_{G}^{(1)}+\nonumber\\&&(-2r^9+12r^8M+16r^6M^3-24r^7M^2)
\partial_{r^2}G^{(1)}\partial_{r}K_{K}^{(1)}+\nonumber\\&&
(8r^7+216r^5M^2-68r^6M+160r^3M^4-304r^4M^3)\partial_{r^2}h_{1}^{(1)}
K_{G}^{(1)}+
\nonumber\\&&(14r^8+22r^7M+464r^5M^3-294r^6M^2-152r^4M^4)\partial_{r^2}G^{(1)}
K_{G}^{(1)}+\nonumber\\&&(-4r^8-96r^6M^2+32r^7M-64r^4M^4+128r^5M^3)
\partial_{r^2}G^{(1)}\partial_{r}K_{r\theta}^{(1)}+\nonumber\\&&
(-2r^9-24r^7M^2+12r^8M+16r^6M^3)\partial_{r}K^{(1)}\partial_{r^2}K_{G}^{(1)}+
\nonumber\\&&(2r^{10}-64r^7M^3+32r^6M^4-16r^9M+48r^8M^2)
\partial_{r^2}G^{(1)}\partial_{r^2}K_{G}^{(1)}+\nonumber\\&&
(-2r^8-4r^7M-48r^5M^3+40r^6M^2)\partial_{r^2}K^{(1)}K_{G}^{(1)}+
\nonumber\\&&(-4r^8+32r^7M-64r^4M^4+128r^5M^3-96r^6M^2)
\partial_{r}h_{1}^{(1)}\partial_{r^2}K_{G}^{(1)}+\nonumber\\&&
(-2r^9-24r^7M^2+12r^8M+16r^6M^3)\partial_{r}G^{(1)}\partial_{r^2}K_{K}^{(1)}+
\nonumber\\&&
(2r^9-16r^6M^3-12r^8M+24r^7M^2)\partial_{r^3}G^{(1)}K_{K}^{(1)}+\nonumber\\&&
(-2r^{10}+64r^7M^3+16r^9M-48r^8M^2-32r^6M^4)
\partial_{r^3}G^{(1)}\partial_{r}K_{G}^{(1)}+\nonumber\\&&
(-112r^6M^3+120r^7M^2-52r^8M+32r^5M^4+8r^9)
\partial_{r}G^{(1)}\partial_{r^2}K_{G}^{(1)}+\nonumber\\&&
(6r^7M-8r^5M^3-2r^8)\partial_{r^2}G^{(1)}K_{K}^{(1)}+
\nonumber\\&&(-24r^7M-32r^5M^3+48r^6M^2+4r^8)
\partial_{r^2}G^{(1)}K_{rr}^{(1)}+\nonumber\\&&
(24r^7M+32r^5M^3-48r^6M^2-4r^8)K^{(1)}\partial_{r^2}K_{G}^{(1)}+\nonumber\\&&
(-24r^7M+48r^6M^2-32r^5M^3+4r^8)H_{2}^{(1)}\partial_{r^2}K_{G}^{(1)}+
\nonumber\\&&(48r^5M^2-32r^4M^3-24r^6M+4r^7)h_{1}^{(1)}\partial_{r^2}K_{K}^{(1)}+
\nonumber\\&&
(-48r^5M^2-8r^7-32r^4M^3+40r^6M+64r^3M^4)h_{1}^{(1)}\partial_{r^2}K_{G}^{(1)}+
\nonumber\\&&
(-80r^4M^3+96r^3M^4+36r^6M-8r^7-24r^5M^2)\partial_{r^2}G^{(1)}K_{r\theta}^{(1)}+
\nonumber\\&&(24r^6M-48r^5M^2+32r^4M^3-4r^7)\partial_{r^2}h_{1}^{(1)}K_{K}^{(1)}+
\nonumber\\&&
(96r^6M^3-48r^7M^2+8r^8M-64r^5M^4)\partial_{r^2}G^{(1)}\partial_{r}K_{G}^{(1)}+
\nonumber\\&&
(56r^5M^3+30r^7M-4r^8-72r^6M^2)\partial_{r}G^{(1)}\partial_{r}K_{K}^{(1)}+
\nonumber\\&&
(24r^6M-6r^7-24r^5M^2)G^{(1)}\partial_{r}K_{K}^{(1)}+
\nonumber\\&&(-144r^4M^3-14r^7+104r^5M^2+12r^6M)\partial_{r}K^{(1)}K_{G}^{(1)}+
\nonumber\\&&
(-144r^5M-32r^3M^3+120r^4M^2+96r^2M^4+40r^6)h_{1}^{(1)}\partial_{r}K_{G}^{(1)}+
\nonumber\\&&
(-16r^5M+64r^4M^2-64r^3M^3)\partial_{r}K^{(1)}K_{r\theta}^{(1)}+
\nonumber\\&&(-24r^4M^2+24r^5M-6r^6)G^{(1)}K_{rr}^{(1)}+\nonumber\\&&
(24r^4M^3-28r^5M^2-4r^7+16r^6M)\partial_{r}G^{(1)}K_{K}^{(1)}+
\nonumber\\&&
(504r^4M^3+58r^7-64r^6M-356r^5M^2)\partial_{r}G^{(1)}K_{G}^{(1)}+
\nonumber\\&&(96r^4M^2-48r^5M-64r^3M^3+8r^6)K^{(1)}\partial_{r}K_{r\theta}^{(1)}+
(300r^4M^2+42r^5M-96r^6)G^{(1)}K_{G}^{(1)}+\nonumber\\&&
(12r^5+96r^3M^2-48r^2M^3-60r^4M)h_{1}^{(1)}K_{rr}^{(1)}+
\nonumber\\&&(16r^5M^3-24r^6M^2+12r^7M-2r^8)
\partial_{r}H_{2}^{(1)}\partial_{r}K_{G}^{(1)}+\nonumber\\&&
(136r^4M+112r^2M^3-24r^5-232r^3M^2)h_{1}^{(1)}K_{K}^{(1)}+\nonumber\\&&
(-96r^4M^2-8r^6+64r^3M^3+48r^5M)H_{2}^{(1)}\partial_{r}K_{r\theta}^{(1)}+
\nonumber\\&&(192r^4M^2-64r^5M-256r^3M^3+128r^2M^4+8r^6)
\partial_{r}h_{1}^{(1)}\partial_{r}K_{r\theta}^{(1)}+\nonumber\\&&
(8r^5M^3+8r^6M^2+4r^8-14r^7M)\partial_{r}K^{(1)}\partial_{r}K_{G}^{(1)}+\nonumber\\&&
(-2r^7+8r^5M^2-16r^4M^3+4r^6M)H_{2}^{(1)}\partial_{r}K_{G}^{(1)}+\nonumber\\&&
(-16r^5M^3-12r^7M+24r^6M^2+2r^8)\partial_{r}G^{(1)}\partial_{r}K_{rr}^{(1)}+
\nonumber\\&&(16r^5M+256r^3M^3+64r^2M^4-240r^4M^2+16r^6)
\partial_{r}G^{(1)}K_{r\theta}^{(1)}+\nonumber\\&&
(-144r^4M+36r^5+144r^3M^2)G^{(1)}K_{r\theta}^{(1)}+\nonumber\\&&
(16r^4M^3+20r^6M-4r^7-32r^5M^2)K^{(1)}\partial_{r}K_{G}^{(1)}+
\nonumber\\&&(-72r^4M+16r^5+48r^3M^2+160r^2M^3-192rM^4)
\partial_{r}h_{1}^{(1)}K_{r\theta}^{(1)}+\nonumber\\&&
(-976r^2M^3+928r^3M^2+12r^5+352rM^4-288r^4M)h_{1}^{(1)}K_{G}^{(1)}+\nonumber\\&&
(64r^2M^3-80r^4M+16r^5+96r^3M^2-128rM^4)h_{1}^{(1)}\partial_{r}K_{r\theta}^{(1)}+
\nonumber\\&&(64r^2M^3-24r^5-160r^3M^2+112r^4M)K^{(1)}K_{r\theta}^{(1)}+
(12r^5+112r^3M^2-64r^2M^3-64r^4M)H_{2}^{(1)}K_{r\theta}^{(1)}+\nonumber\\&&
(-88r^4+256r^3M+96r^2M^2-640rM^3+256M^4)h_{1}^{(1)}K_{r\theta}^{(1)}+
\nonumber\\&&(44r^6+18r^5M-212r^4M^2)K^{(1)}K_{G}^{(1)}+
(-6r^7+24r^6M-24r^5M^2)G^{(1)}\partial_{r}K_{G}^{(1)}+\nonumber\\&&
(36r^6-24r^5M-96r^4M^2)G^{(1)}K_{K}^{(1)}+\nonumber\\&&
(-32r^4M^3+4r^7+48r^5M^2-24r^6M)\partial_{r}h_{1}^{(1)}
\partial_{r}K_{K}^{(1)}+\nonumber\\&&(-64r^3M^4-240r^5M^2+224r^4M^3-16r^7+104r^6M)
\partial_{r}G^{(1)}\partial_{r}K_{r\theta}^{(1)}+\nonumber\\&&
(288r^3M^3-276r^4M^2-336r^2M^4+228r^5M-60r^6)
\partial_{r}h_{1}^{(1)}K_{G}^{(1)}+
\nonumber\\&&(-32r^4M^3+48r^5M^2+4r^7-24r^6M)
\partial_{r}K^{(1)}\partial_{r}K_{r\theta}^{(1)}+\nonumber\\&&
(88r^4M^3-96r^5M^2+30r^6M-2r^7)\partial_{r}H_{2}^{(1)}K_{G}^{(1)}+\nonumber\\&&
(-48r^4M^2+24r^5M-4r^6+32r^3M^3)\partial_{r}H_{2}^{(1)}K_{r\theta}^{(1)}+
\nonumber\\&&(-40r^4M^3+48r^5M^2-18r^6M+2r^7)\partial_{r}G^{(1)}K_{rr}^{(1)}+
\nonumber\\&&
(192r^4M^2+20r^6-108r^5M-112r^3M^3)\partial_{r}h_{1}^{(1)}K_{K}^{(1)}+\nonumber\\&&
(48r^4M^2-12r^5M-48r^3M^3)h_{1}^{(1)}\partial_{r}K_{K}^{(1)}+\nonumber\\&&
(28r^6-12r^5M-88r^4M^2)H_{2}^{(1)}K_{G}^{(1)}+
\nonumber\\&&(48r^5M+64r^3M^3-96r^4M^2-8r^6)
\partial_{r}h_{1}^{(1)}K_{rr}^{(1)}+\nonumber\\&&
(-176r^5M^3+132r^6M^2+16r^4M^4-8r^7M-8r^8)
\partial_{r}G^{(1)}\partial_{r}K_{G}^{(1)}+\nonumber\\&&
(-16r^7+112r^6M-288r^5M^2+320r^4M^3-128r^3M^4)
\partial_{r}h_{1}^{(1)}\partial_{r}K_{G}^{(1)}+
\nonumber\\&&(32r^3M^3+24r^5M-48r^4M^2-4r^6)h_{1}^{(1)}
\partial_{r}K_{rr}^{(1)}+(-6r^6+12r^5M)H_{2}^{(1)}K_{K}^{(1)}+\nonumber\\&&
\left.(-8r^6M+2r^7+8r^5M^2)\partial_{r}H_{2}^{(1)}K_{K}^{(1)}\right]\ .
\label{TDWF}
\end{eqnarray}

\section{Energy radiated} \label{sec:energy} 

Radiation is most clearly analyzed in a reference system that is asymptotically
flat (AF). By this we shall mean a coordinate system in which 
the deviations of the metric $\delta g_{\mu\nu}$ from Minkowski form
decrease with $r$, at constant $t-r^{*}$, according to:
\begin{displaymath}
\delta g_{tt}, \delta g_{tr}, \delta g_{rr}\sim{\cal
O}(r^{-3})\quad\quad \delta g_{tr}, \delta g_{r\theta}\sim{\cal
O}(r^{-1})\quad\quad 
\end{displaymath}
\begin{equation}\label{AFconds}
\delta g_{\theta\theta}+\delta g_{\theta\phi}/\sin^{2}\theta\sim{\cal O}(r^{1})\quad\quad
\delta g_{\theta\phi},
\delta g_{\theta\theta}-\delta g_{\theta\phi}/\sin^{2}\theta\sim{\cal O}(r^{3})\ .
\end{equation}
In such a coordinate system, the power carried by gravitational waves is
given by\cite{LL,CPMII}
\begin{equation}\label{LLformula}
{d{\rm Power} \over d\Omega}= {1\over 16\pi r^2}
\left[
\frac{1}{\sin^{2}{\theta}}
\left(
\frac{\partial
g_{{\theta}{\phi}}}{\partial t}
\right)^2 +
\frac{1}{4} 
\left(
\frac{\partial g_{{\theta}{\theta}}}{\partial t} - 
\frac{1}{\sin^{2}{\theta}}\frac{\partial g_{{\phi}{\phi}}}{\partial t}
\right)^2
\right]
\end{equation}
In the statement of the conditions for an AF gauge, and in
(\ref{LLformula}), there is no reference to the order of the metric
perturbations in some expansion parameter $\epsilon$. To compute the
power to first order in $\epsilon$, one finds the perturbations
$\delta g_{{\theta}{\theta}}, \delta
g_{{\theta}{\phi}}, \delta g_{{\phi}{\phi}}$ in a
gauge that satisfies the AF conditions to first order in $\epsilon$
and uses those values in (\ref{LLformula}). For a computation correct
to second order, the values of $\delta g_{{\theta}{\theta}},
\delta g_{{\theta}{\phi}}, \delta g_{{\phi}{\phi}}$
used in (\ref{LLformula}) must in principle be computed in a gauge
that satisfies the AF conditions to second order in $\epsilon$.

In practice, the use of gauge invariant quantities $\Psi^{(1)}$ and
$\Psi^{(2)}_{\rm RW}$ in our computations simplifies the evaluation of
energy. A first and second order gauge transformation can always be
done to bring the metric perturbations into a form that is AF to first
and second order, and $\Psi^{(1)}$ and $\Psi^{(2)}_{\rm RW}$ are unaffected by
such a transformation. We can, therefore, treat them as if they had
been computed in a gauge that is first and second order AF, and we
need only read off the first and second order metric perturbations
from $\Psi^{(1)}$ and $\Psi^{(2)}_{\rm RW}$. When restricted to first order
only, this is the method that was used to compute radiated power
in Refs.\,\cite{PP} and \cite{CPMII}.

If we take the expressions in (\ref{METRICpenult}) and
(\ref{METRICend}) to be in the AF gauge, then for our axisymmetric
quadrupole example the expression in (\ref{LLformula}) becomes
\begin{equation}\label{LLformulaG}
{d{\rm Power} \over d\Omega}= {r^{2}\over 64\pi}
\left(
\epsilon\frac{\partial G^{(1)}}{\partial t}
+\frac{\epsilon^{2
}}{2}\frac{\partial G^{(2)}}{\partial t}
\right)^2 \left(
P_{2}^{''}-\cot\theta 
P_{2}^{'}
\right)^{2 }
={9r^{2}\over 64\pi}
\left(
\epsilon\frac{\partial G^{(1)}}{\partial t}
+\frac{\epsilon^{2
}}{2}\frac{\partial G^{(2)}}{\partial t}
\right)^2 \sin^{4}\theta\ .
\end{equation}
The first order part is easily dealt with, and the treatment is identical to
that in equations (III-21) -- (III-26) of Ref.\,\cite{CPMII}.
From the definition of $
\Psi^{(1)}$ in
(\ref{WFFO})
and the conditions in (\ref{AFconds})
it follows that in the AF gauge
\begin{equation}
12rG^{(1)}=
\Psi^{(1)}+{\cal O}(r^{-1})\ .
\end{equation}
Evaluating the second order part is considerably more difficult.
Using the AF conditions in (\ref{AFconds}), one must solve
(\ref{def:L2}) for $G^{(2)}$. The result will contain terms that 
are linear in second order perturbations and quadratic in first
order perturbations. The linear terms are identical to those in the 
first order analysis, so the result can be written in the form
\begin{equation}\label{G2andPsi2}
12rG^{(2)}=
\Psi^{(2)}_{\rm RW}-Q^{(1)}_{\rm RW}
+{\cal O}(r^{-1})\ .
\end{equation}
Here $Q^{(1)}_{\rm RW}$ is the set of terms quadratic in first order
perturbations, as defined in (\ref{eq:defpsi2}), and as explicitly
exhibited in (\ref{def:Q1}). It remains to find $Q^{(1)}_{\rm RW}$ to ${ \cal
O}(r^{0 })$ in the AF gauge. The result must be expressible entirely
in terms of $\Psi^{(1)}$, since (\ref{LLformulaG}) is
an expression for a physical quantity, and $\Psi^{(1)}$ contains all
the gauge invariant first order information about perturbations.

To find $Q^{(1)}_{\rm RW}$ in the AF gauge, we start by writing the following
asymptotic expansions for first order perturbative quantities:
\begin{eqnarray}
H_{0}^{(1)}& =&f_{0}/r^{3}+\cdots\nonumber\\
H_{1}^{(1)}& =&f_{1}/r^{3}+\cdots\nonumber\\
H_{2}^{(1)}& =&f_{2}/r^{3}+\cdots\nonumber\\
h_{0}^{(1)}& =&f_{3}/r+I_{0}/r^{2}+\cdots\nonumber\\
h_{1}^{(1)}& =&f_{4}/r+I_{1}/r^{2}+\cdots\nonumber\\
K^{(1)}& =&f_{5}/r+f_{6}/r^{2}+ f_{7}/r^{3}+\cdots\nonumber\\
G^{(1)}& =&f_{8}/r+f_{9}/r^{2}+ f_{10}/r^{3}+\cdots\label{AFexpansion} ,
\end{eqnarray}
where $f_{1}\cdots f_{10},I_{0},I_{1}$ 
are functions of $t-r^{*}$.
When (\ref{AFexpansion}) is used in the expression for $Q^{(1)}_{\rm RW}$
given in
(\ref{def:Q1}), the result is an expression of the form
\begin{equation}
Q^{(1)}_{\rm RW}={\cal Q}_{0}+r{\cal Q}_{1}+r^{2}{\cal Q}_{2}+{\cal O}(r^{-1})\ ,
\end{equation}
where ${\cal Q}_{0}, {\cal Q}_{1}, {\cal Q}_{2}$ are functions of
$t-r^{*}$. 
As an example, we present here the
 explicit expression for ${\cal Q}_{2}$:
\begin{equation}
{\cal Q}_{2}=-2f_{8}'f_{5}''+10f_{8}'f_{8}''\ ,
\end{equation}
where a prime ($'$) denotes differentiation with respect to $t-r_{*}$.
The expressions for ${\cal Q}_{0}$ and ${\cal Q}_{1}$ are very lengthy
and will not be explicitly exhibited.

The radiated power is a physical quantity. Since $\Psi^{(1)}$ and
$\Psi^{(2)}_{\rm RW}$ carry all the (first and second order) gauge
invariant information about the spacetime, it must be possible to
express the coefficients ${\cal Q}_{0}, {\cal Q}_{1}, {\cal Q}_{2}$ in
terms of $\Psi^{(1)}$. To find such expressions we start by using the
AF expansions of (\ref{AFexpansion}) in Einstein's vacuum equations to
derive
\begin{equation}\label{AFG1}
3f_{8}'-f_{5}'=0
\end{equation}
\begin{equation}\label{AFG2}
6Mf_{4}'-3f_{10}'+7Mf_{5}-2f_{6}-3I_{1}'+f_{7}'
-9Mf_{8}-3f_{3}-3I_{0}' =0
\end{equation}
\begin{equation}\label{AFG3}
f_{5}-f_{8}-f_{9}'=0
\end{equation}
\begin{equation}\label{AFG4}
f_{3}'+f_{4}'+f_{9}'=0
\end{equation}
\begin{equation}\label{AFG5}
4f_{10}'-4Mf_{4}'+2I_{1}'+2f_{9}-2Mf_{8}-4Mf_{9}'+2f_{4}+2I_{0}'=0
\end{equation}
\begin{equation}\label{AFG6}
f_{6}''-3f_{9}''=0
\end{equation}
\begin{equation}\label{AFG7}
2Mf_{5}'-f_{6}'+2f_{7}''-9Mf_{8}'\ .
\end{equation}
The first two equations are respectively the $r^{6}$ and $r^{4}$ parts
of the Einstein equation $G_{tt}+G_{tr}=0$;
Eqs.\,(\ref{AFG3}), (\ref{AFG4}), (\ref{AFG5}) are combinations of the
$r^{4}$ and $r^{5}$ parts of $G_{\theta\theta}=0$ and
$G_{\phi\phi}=0$; Eq.\,(\ref{AFG6}) is the $r^{5}$ part of
$G_{rr}=0$; Eq.\,(\ref{AFG7}) is a combination of the $r^{3}$ parts of
$G_{\theta\theta}=0$ and $G_{\theta\theta}=0$, and the $r^{5}$ part of
$G_{rr}=0$.  We use (\ref{AFG4})  in the form
$f_{3}+f_{4}+f_{9}=0$. The justification for this is that $f_{3}$,
$f_{4}$ and $f_{9}$ must be zero in a stationary
solution\cite{KipRMP}. The integration constant must be zero
therefore, when (\ref{AFG4}) is integrated. Similar arguments justify
integration of (\ref{AFG6}). With (\ref{AFG1}) --(\ref{AFG7}), and 
considerable manipulation, we end up with 
\begin{equation}\label{Q0}
{\cal Q}_{0}=\frac{1}{6048}\left(
36\Psi^{(1)}\partial_{t}\Psi^{(1)}-19M\partial_{t}\Psi^{(1)}\partial_{t}\Psi^{(1)}\right)\nonumber
\end{equation}\label{Q1}
\begin{equation}
{\cal Q}_{1}=\frac{1}{1008}
\left(
7\,\partial_{t}\Psi^{(1)}\partial_{t}\Psi^{(1)}-4M\,\Psi^{(1)}\partial_{tt}\Psi^{(1)}\right)
\end{equation}
\begin{equation}\label{Q2}
{\cal Q}_{2}=\frac{1}{126}
\,\partial_{t}\Psi^{(1)}\partial_{tt}\Psi^{(1)}\ .
\end{equation}

Notice that ${\cal Q}_{1}$ and ${\cal Q}_{2}$ do not vanish, and hence
$Q^{(1)}_{\rm RW}$ diverges at $r\rightarrow\infty$.  It is important
to understand that this is not incompatible with with asymptotic
flatness. The value of $G^{(2)}$ must fall off as $r^{-1}$ in AF
coordinates, so $Q^{(1)}_{\rm RW}$ can diverge if there is a
compensating divergence in $\Psi^{(2)}_{\rm RW}$.  There must, in fact,
be a divergence of this order, since the source term in
(\ref{eq:PsiRW}) turns out to diverge. In practice, numerical
computations with divergent quantities are to be avoided. It is
useful, therefore, to exploit the fact that the second order wave
function is not unique. (See the discussion at the end of
Sec.\,\ref{sec:intro}.) We now introduce an alternative second
order wave function $\Psi^{(2)}_{\rm rad}$ by
\begin{equation}\label{def:Psi2rad}
\Psi^{(2)}_{\rm rad}
=\Psi^{(2)}_{\rm RW}+\Xi_{\rm rad}\ ,
\end{equation}
where
\begin{displaymath}
\Xi_{\rm rad}\equiv
-\frac{1}{2016}\left(
144\Psi^{(1)}\partial_{t}\Psi^{(1)}
-76M\partial_{t}\Psi^{(1)}\partial_{t}\Psi^{(1)}\right.\nonumber
\end{displaymath}
\begin{equation}\label{def:Xirad}
\left.+
r\left[56\,\partial_{t}\Psi^{(1)}\partial_{t}\Psi^{(1)}-32M\,\Psi^{(1)}\partial_{tt}\Psi^{(1)}
\right]+ 16\,r^{2}\partial_{t}\Psi^{(1)}\partial_{tt}\Psi^{(1)}
\right)\ .
\end{equation}
The wave equation for $\Psi^{(2)}_{\rm rad}$,
\begin{equation}\label{eq:Psirad}
\widehat{Z}(\Psi^{(2)}_{\rm rad})=
{\cal S}_{\rm RW}+\widehat{Z}(\Xi_{\rm rad})\equiv{\cal S}_{\rm rad}
\ ,
\end{equation}
has a source term that is well behaved at $r\rightarrow\infty$, and
in AF coordinates, we then have from (\ref{G2andPsi2}) and
(\ref{Q0})-- (\ref{def:Xirad}), that
\begin{equation}\label{G2andPsi2rad}
12rG^{(2)}=
\Psi^{(2)}_{\rm rad}+{\cal O}(r^{-1})\ ,
\end{equation}
and thus
\begin{equation}\label{LLformula2}
{d{\rm Power} \over d\Omega}= {1\over 1024\pi}
\left(
\epsilon\frac{\partial \Psi^{(1)}}{\partial t}
+\frac{\epsilon^{2
}}{2}\frac{\partial \Psi^{(2)}_{\rm rad}}{\partial t}
\right)^2\sin^{4}\theta\ .
\end{equation}
Integration over all angles then gives us the total power
\begin{equation}
{\rm Power}= {1\over 480}
\left(
\epsilon\frac{\partial \Psi^{(1)}}{\partial t}
+\frac{\epsilon^{2
}}{2}\frac{\partial \Psi^{(2)}_{\rm rad}}{\partial t}
\right)^2\ .
\end{equation}

\section{Summary and Discussion}
\label{sec:disc}

For convenience, we repeat and summarize here the main results of the
paper.  The wavefunction
\begin{eqnarray}
\Psi^{(1)} &=&{r \over 6\,(2r+3M)}\left[2 (r-2M) (H_{2}^{(1)}-r \partial_r K^{(1)})-2 (r-3M)
K^{(1)} + 6 \left\{r K^{(1)}+ {(r-2M) \over  r} (r^2 \partial_r G^{(1)}-2 h_{1}^{(1)})\right\}\right]\ ,
\end{eqnarray}
is Moncrief's\cite{MON} wavefunction. It is constructed completely
from first order perturbations of the 3-geometry on a hypersurface,
and its time derivative can be found from the first order
perturbations of the 3-geometry and extrinsic curvature of the
hypersurface.  This wave function is invariant with respect to first
order gauge transformations, and satisfies the Zerilli equation
\begin{equation}
\left[-{\partial^2 \over \partial t^2} 
+{\partial^2 \over \partial r_*^2} +V(r)\right]
\Psi^{(1)} = 0\ ,
\end{equation}
where $V(r)$ is given in (\ref{def:V}).  From (\ref{def:L2}),
(\ref{def:Psi2rad}),(\ref{def:Xirad}), our second order wave function is
\begin{equation}
\Psi_{\rm rad
}^{(2)}={r \over 6\,(2r+3M)}\left[2 (r-2M) (H_{2}^{(2)}-r \partial_r
K^{(2)})-2 (r-3M) K^{(2)}+\right. 
\end{equation}
\begin{displaymath}
\left. 6 \left\{r K^{(2)}+ {(r-2M) \over r} (r^2
\partial_r G^{(2)}-2 h_{1}^{(2)})\right\}\right]
-\frac{1}{2016}\left(
144\Psi^{(1)}\partial_{t}\Psi^{(1)}
-76M\partial_{t}\Psi^{(1)}\partial_{t}\Psi^{(1)}\right.\nonumber
\end{displaymath}
\begin{equation}
\left.+
r\left[56\,\partial_{t}\Psi^{(1)}\partial_{t}\Psi^{(1)}-32M\,\Psi^{(1)}\partial_{tt}\Psi^{(1)}
\right]+ 16\,r^{2}\partial_{t}\Psi^{(1)}\partial_{tt}\Psi^{(1)}
\right)+Q^{(1)}_{\rm RW}(\Psi^{(1)},\Psi^{(1)})\ ,
\end{equation}
where the explicit form of $Q^{(1)}_{\rm RW}$ is given in
(\ref{def:Q1}).  The wave function $\Psi_{\rm rad}^{(2)}$ gives the
second order equivalent of the two important advantages of the
Moncrief function $\Psi^{(1)}$: (i) it is gauge invariant, as spelled
out in Sec.\, \ref{sec:1and2inv} and (ii) its value and its time
derivative can be found directly from the first and second order
perturbations of the 3-geometry and extrinsic curvature of a
hypersurface, as described in Sec.\,\ref{sec:id}.  The second order
wave function satisfies a Zerilli equation
\begin{equation}
\left[-{\partial^2 \over \partial t^2} 
+{\partial^2 \over \partial r_*^2} +V(r)\right]
\Psi^{(2)}_{\rm rad}= {\cal S}_{\rm rad}\ .
\end{equation}
The source term ${\cal S}_{\rm rad}$ is well behaved at infinity, and
is given by (\ref{eq:PsiRW}), (\ref{def:Xirad}) and (\ref{eq:Psirad}). 
The outgoing solution for  $\Psi^{(2)}_{\rm rad}$ has the asymptotic 
behavior of a function of retarded time $t-r^{*}$, and the gravitational
wave power carried in the perturbations is given by the simple 
prescription in (\ref{LLformula2}).

The scheme outlined in the above paragraph gives a formalism that is
definitive and complete (for the case of even parity, axisymmetric,
quadrupole perturbations). This formalism involves only gauge
invariant variables, and requires the solution of a wave equation with
a source that is well behaved at spatial infinity. Due to the gauge
invariant nature of the wave function, the Cauchy data for the wave
equation can be constructed immediately from an initial value solution
given in any gauge.

The method given here constitues a reformulation  of the fixed-gauge approach
used by Gleiser {\it et
al.}\cite{GNPP1,GNPP2,GNPP3,G_masscrrct,GNPPBY,GNPPboost}. The gauge invariant
approach appears to offer advantages both in the organization  of 
computations, and in the conceptual clarity of the gauge invariant variables.

\acknowledgments We thank William Krivan, Carlos Lousto, Jorge Pullin,
and Reinaldo Gleiser for useful discussions. We gratefully acknowledge
the support of the National Science Foundation under grant PHY9734871.

\end{document}